\documentclass[pre,aps,floats,epsf,color,twocolumn]{revtex4}
\usepackage{graphicx}
\usepackage{dcolumn}
\usepackage{bm}
\begin{document}

\title
{The Effect of Annealing Temperature on Statistical Properties of
$WO_3$ Surface}
\author
{ G. R. Jafari $^{a,b}$, A. A. Saberi $^c$, R. Azimirad $^c$, A. Z.
Moshfegh $^c$, and  S. Rouhani $^c$}
\address
{\it $^a$ Department of Physics, Shahid Beheshti University, Evin,
Tehran 19839, Iran \\ $^b$ Department of Nano-Science, IPM, P. O.
Box 19395-5531, Tehran, Iran  \\ $^c$ Department of Physics, Sharif
University of Technology, P.O. Box 11365-9161, Tehran, Iran }

\begin{abstract}

We have studied the effect of annealing temperature on the
statistical properties of $WO_3$ surface using atomic force
microscopy techniques (AFM). We have applied both level crossing and
structure function methods. Level crossing analysis indicates an
optimum annealing temperature of around 400$^oC$ at which the
effective area of the $WO_3$ thin film is maximum, whereas
composition of the surface remains stoichiometric. The complexity of
the height fluctuation of surfaces was characterized by roughness,
roughness exponent and lateral size of surface features. We have
found that there is a phase transition at around 400$^oC$ from one
set to two sets of roughness parameters. This happens due to
microstructural changes from amorphous to crystalline structure in
the samples that has been already found experimentally.


\end{abstract}

\maketitle
 \hspace{.3in}

\section{Introduction}

Transition metal oxides represent a large family of materials
possessing various interesting properties, such as
superconductivity, colossal magneto-resistance and
piezoelectricity. Among them, tungsten oxide is of intense
interest and has been investigated extensively for its
distinctive properties. With outstanding electrochromic
\cite{Granqvist,Bueno,Azimirad,Siokou,Kuai,Takeda}, photochromic
\cite{Avellaneda}, gaschromic \cite{Lee}, gas sensor
\cite{Kim,György,Kawasaki}, photo-catalyst  \cite{Gondal}, and
photoluminescence properties \cite{Feng}, tungsten oxide has been
used to construct "smart-window", anti-glare rear view mirrors of
automobile, non-emissive displays, optical recording devices,
solid-state gas sensors, humidity and temperature sensors,
biosensors, photonic crystals, and so forth.

$WO_3$ thin films can be prepared by various deposition techniques
such as thermal evaporation \cite{Azimirad,Lee}, spray pyrolysis
\cite{Hao}, sputtering \cite{Takeda}, pulsed laser ablation
\cite{György,Kawasaki}, sol-gel coating \cite{Bueno,Kuai,Garcia},
and chemical vapor deposition \cite{Seman}.

The gas sensitivity of $WO_3$ heavily depends upon film parameters
such as composition, morphology (e.g. grain size), nanostructure and
microstructure (e.g. porosity, surface-to-volume ratio). Film
parameters are related to the deposition technique used, the
deposition conditions and the subsequent annealing process.
Annealing, which is an essential process to obtain stable films with
well-defined microstructure, causes stoichiometry and
microstructural changes that have a high influence on the sensing
characteristics of the films \cite{Stankova}. Moreover, the surface
structure and surface morphology of the metal oxides are also
important for different applications. In fact, the electrochromic
devices are made of amorphous oxides \cite{Granqvist}, while
crystalline phase plays a major role in catalysts and sensors
\cite{Stankova}. This is because, the minor change in their chemical
composition and crystalline structure
could modify different properties of the metal oxides.\\

In practice, one of the effective ways to modify the surface
morphology is annealing process at various temperatures. So far,
most of morphological analysis related to the $WO_3$ surface were
accessible through the experimental methods. Usually these analysis
are rigorous and time consuming. Moreover, lack of the suitable
analysis for AFM data to find the nano and microstructural
properties of surfaces was feeling perfectly.\\

In this article, we introduce the methods: roughness analysis and
level crossing as suitable candidates and show that we can get
easily the structure and morphological properties of a surface in
a fast manner, only using the AFM observation as an initial data.\\

The roughness of a surface has been studied as a simple growth model
using analytical and numerical methods
\cite{Barabasi,Jafari10,Jafari11,Irajizad,Halpin,Krug,Meakin,Kardar,Masoudi}.
These studies quite generally proposed that the height fluctuations
have a self-similar character and their average correlations exhibit
a dynamic scaling form. Also some authors recently use the average
frequency of positive slope level crossing to provide further
complete analysis on roughness of a surface \cite{Tabar}. This
stochastic approach has turned out to be a promising tool also for
other systems with scale dependent complexity, such as in surface
growth where one would like to measure the roughness \cite{Tabar1}.
Some authors have applied this method to study the fluctuations of
velocity fields in Burgers turbulence \cite{Movahed} and the
Kardar-Parisi-Zhang equation in (d+1)-dimensions \cite{Bahraminasab}
and analyzing the stock market \cite{Jafari2}.\\

In this work, we have used the scaling analysis to determine the
roughness, roughness exponent and the lateral size of surface
features. Moreover, level crossing analysis has been
utilized to estimate the effective area of a surface.\\

This paper is organized as follows: In section II, we have discussed
about the film preparation and experimental results obtained from
AFM, XPS and UV-visible spectrophotometer for the annealed samples
at the various temperatures. In section III, we have introduced the
analytical methods briefly. Data description and data analysis based
on the statistical parameters of $WO_3$ surface as a function of
annealing temperatures are given in section IV.Finally, section V
concludes presented results.


\section{Experimentals}

Thin films of $WO_3$ were deposited on microscope slide glass using
thermal evaporation method. The deposition system was evacuated to a
base pressure of $\sim 4\times10^{-3} Pa$. Thickness of the
deposited films was considered about 200 $nm$ measured by the stylus
and optical techniques. More details about the other deposition
parameters of the films are recently reported elsewhere
\cite{Moshfegh}.

To study the effect of annealing temperature on surface structure
and optical properties of the samples, they were annealed at 200,
300, 350, 400, 450, and 500$^oC$ in air for a period of 60 $min$.
Optical transmission and reflection measurements of the deposited
films were performed in a range of 300-1100 $nm$ wavelength using a
Jascow V530 ultraviolet (UV)-visible spectrophotometer with
resolution of 1 $nm$.

X-ray photoelectron spectroscopy (XPS) using a Specs EA 10 Plus
concentric hemispherical analyzer (CHA) with $Al$ $K_\alpha$ anode
at energy of 1486.6 $eV$ was employed to study the atomic
composition and chemical state of the tungsten oxide thin films. The
pressure in the ultra high vacuum surface analysis chamber was less
than $1.0 \times10^{-7} Pa$. All binding energy values were
determined by calibration and fixing the $C(1s)$ line to 285.0 $eV$.
The XPS data analysis and deconvolution were performed by SDP
(version 4.0) software. The nanoscale Surface topography of the
deposited films was investigated by Thermo Microscope Autoprobe
CP-Research atomic force microscopy (AFM) in air with a silicon tip
of 10 $nm$ radius in contact method. The AFM images were recorded
with resolution of about $20$ $nm$ in a scale of $5 \times 5$ $ \mu m$.\\

\subsection{XPS Characterization}

The elemental and chemical characterizations of the films were
performed by XPS. Figure $1a$ shows the $W(4f)$ core level spectra
recorded on the "as deposited" $WO_3$ sample, and the results of its
fitting analysis. To reproduce the experimental data, one doublet
function was used for the $W(4f)$ component. This contains
$W(4f_7/2)$ at 35.6 $eV$ and $W(4f_5/2)$ at 37.8 $eV$ with a
full-width at half-maximum (FWHM) of $1.75 \pm 0.04$ $eV$. The area
ratio of these two peaks is 0.75 which is supported by the
spin-orbit splitting theory of $4f$ levels. Moreover, the structure
was shifted by ~5 $eV$ toward higher energy relative to the metal
state.\\ It is thus clear that the main peaks in our XPS spectrum
attributed to the $W^{6+}$ state on the surface
\cite{Granqvist,Bueno,Crist}. In stoichiometric $WO_3$, the six
valence electrons of the tungsten atom are transferred into the
oxygen p-like bands, which are thus completely filled. In this case,
the tungsten $5d$ valence electrons have no part of their wave
function near the tungsten atom and the remaining electrons in the
tungsten atom experience a stronger Coulomb interaction with the
nucleus than in the case of tungsten atom in a metal, in which the
screening of the nucleus has a component due to the $5d$ valence
electrons. Therefore, the binding energy of the $W(4f)$ level is
larger in $WO_3$ than in metallic tungsten. If an oxygen vacancy
exists, the electronic density near its adjacent $W$ atom increases,
the screening of its nucleus is higher and, thus, the $4f$ level
energy is expected to be at lower binding energy \cite{Granqvist}.

By increasing the annealing temperature it was observed that the
position of $W(4f)$ peak did not obviously change. But for $WO_3$
thin film annealed at $500^oC$ (Fig. 1b), the $W(4f)$ peak moved to
a lower binding energy so that $W(4f_{7/2})$ position was observed
at 35.0 $eV$. This can be related to oxygen vacancy at this
high annealing temperature and formation of $W^{5+}$.\\
\begin{figure}
\includegraphics[width=8.5truecm]{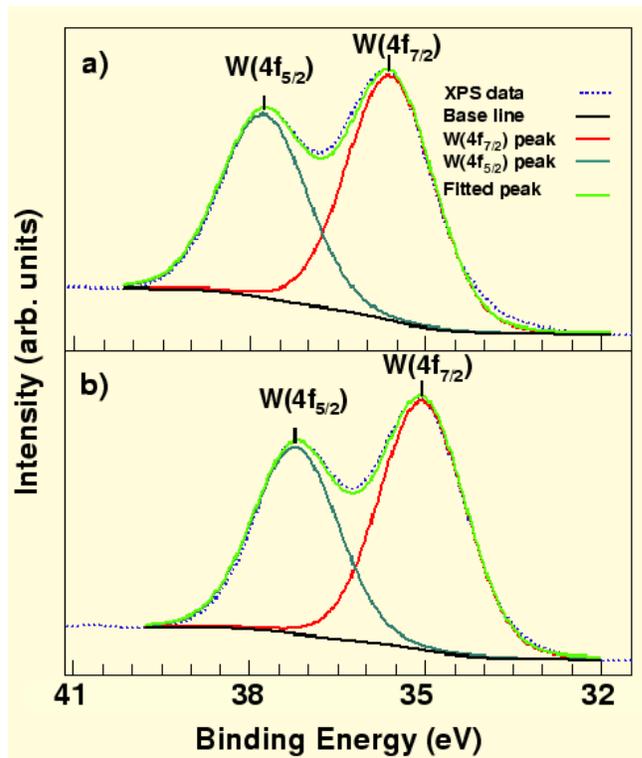}
 \narrowtext \caption{$W(4f)$ core level spectra of $WO_3$ thin films: a)
  "as deposited" and b) annealed at 500$^oC$.}
\end{figure}

\subsection{Optical Characterization}

The transmittance and reflectance spectra in the visible and
infrared range recorded for the $WO_3$ thin films before and after
annealing at different temperatures (Fig. 2a). It is seen that, the
transmittance of the "as deposited" films in the visible range
varies from about 80 up to nearly $100\%$ (without considering the
substrate contribution). Correspondingly, maximum value of the
reflectance for both the film and the substrate is about $20\%$ (the
reflectance from the bare glass substrate was measured about
$10\%$). The sharp reduction in the transmittance spectrum at the
wavelength of $\sim 350 nm$ is due to the fundamental absorption
edge that was also reported previously
\cite{Granqvist,Bueno,Azimirad}.\\ The oscillations in the
transmission and reflection spectra are caused by optical
interference. The optical transmittance of $WO_3$ films strongly
depends on the oxygen content of the films. In fact,
non-stoichiometric films with composition of $WO_{3-x}$ show a blue
tinge for $x > 0.03$ \cite{Lee2}.\\ The "as deposited" pure tungsten
oxide films were highly transparent with no observable blue
coloration, under our experimental conditions. As can be seen from
Fig. 2a after annealing process at 200 to 400$^oC$, the
transmittance and reflectance of the $WO_3$ films have not changed
significantly. Only, the position of the oscillations altered due to
thickness reduction and film condensation after the heat-treatment
process \cite{Granqvist}. At $500^oC$ transmittance and reflectance
of the annealed $WO_3$ film  is reduced about $10\%$, therefore at
this temperature, the film turn into non-stoichiometric composition,
so that it could be seen from
changing color of the film.\\

\begin{figure}
\includegraphics[width=8.5truecm]{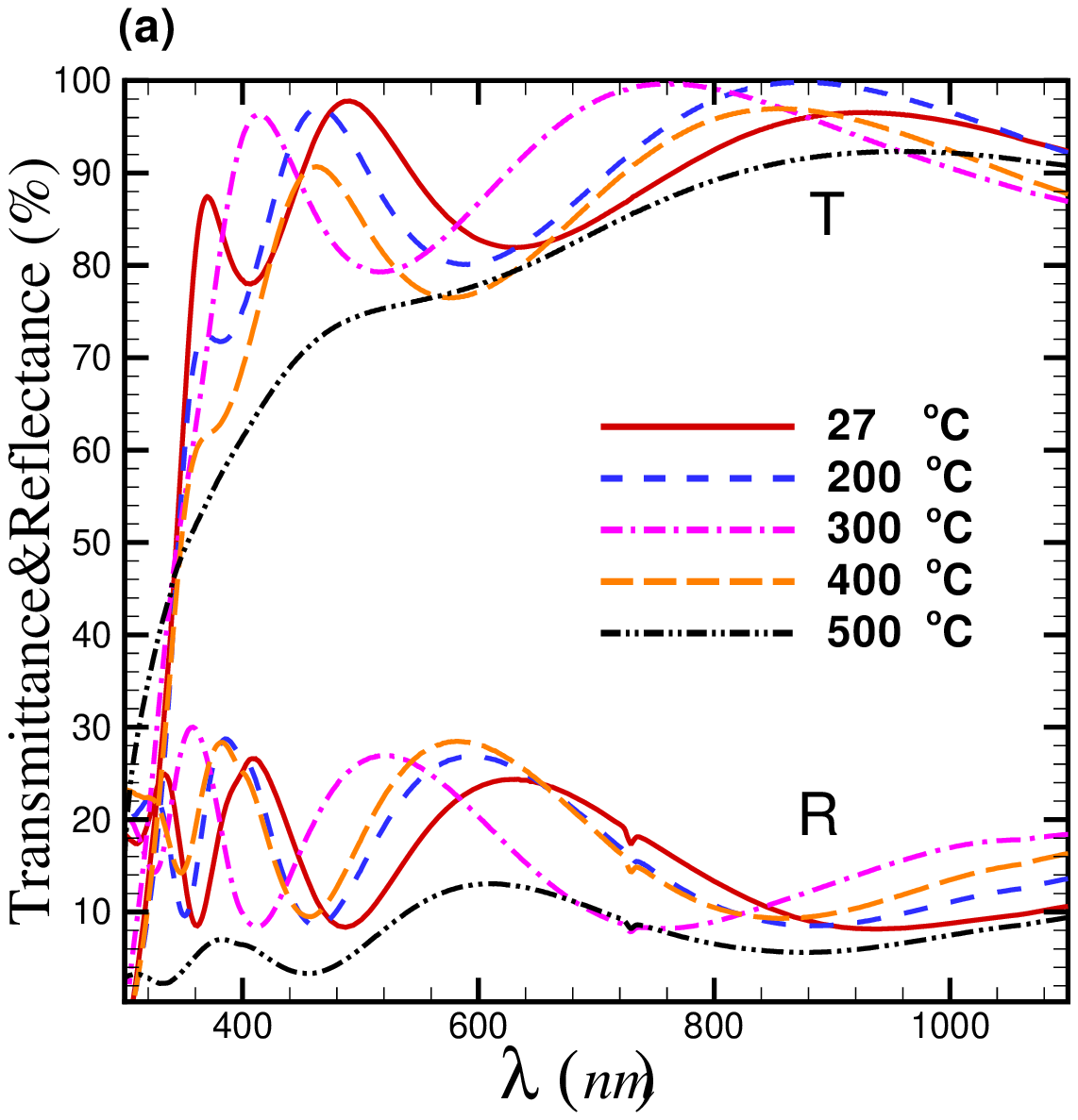}
\includegraphics[width=8.5truecm]{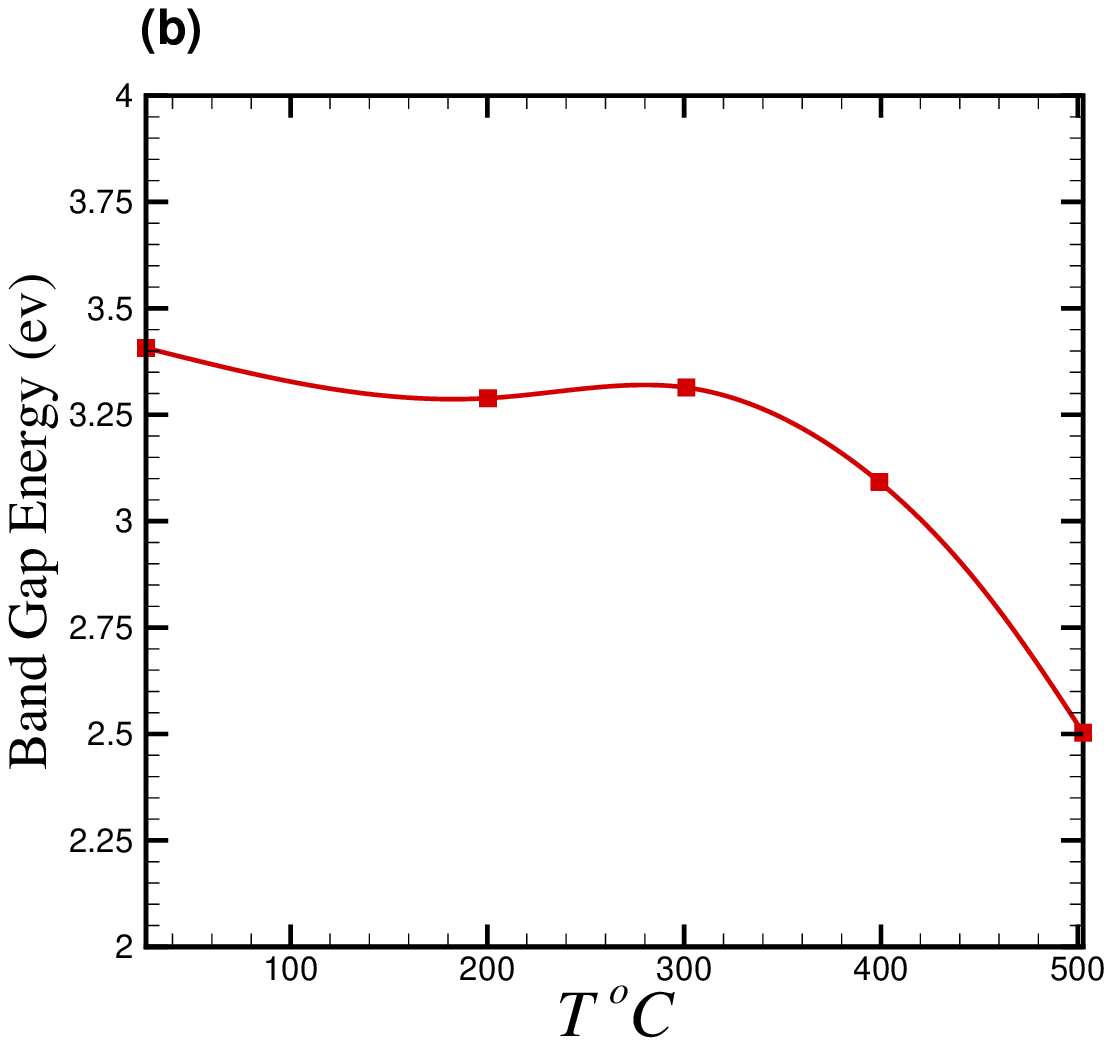}
 \narrowtext \caption{a) Optical transmittance (T) and reflectance (R) and
  b) Optical band gap energy of the $WO_3$ thin films annealed at different temperatures.}
\end{figure}
\begin{figure}
\includegraphics[width=8truecm]{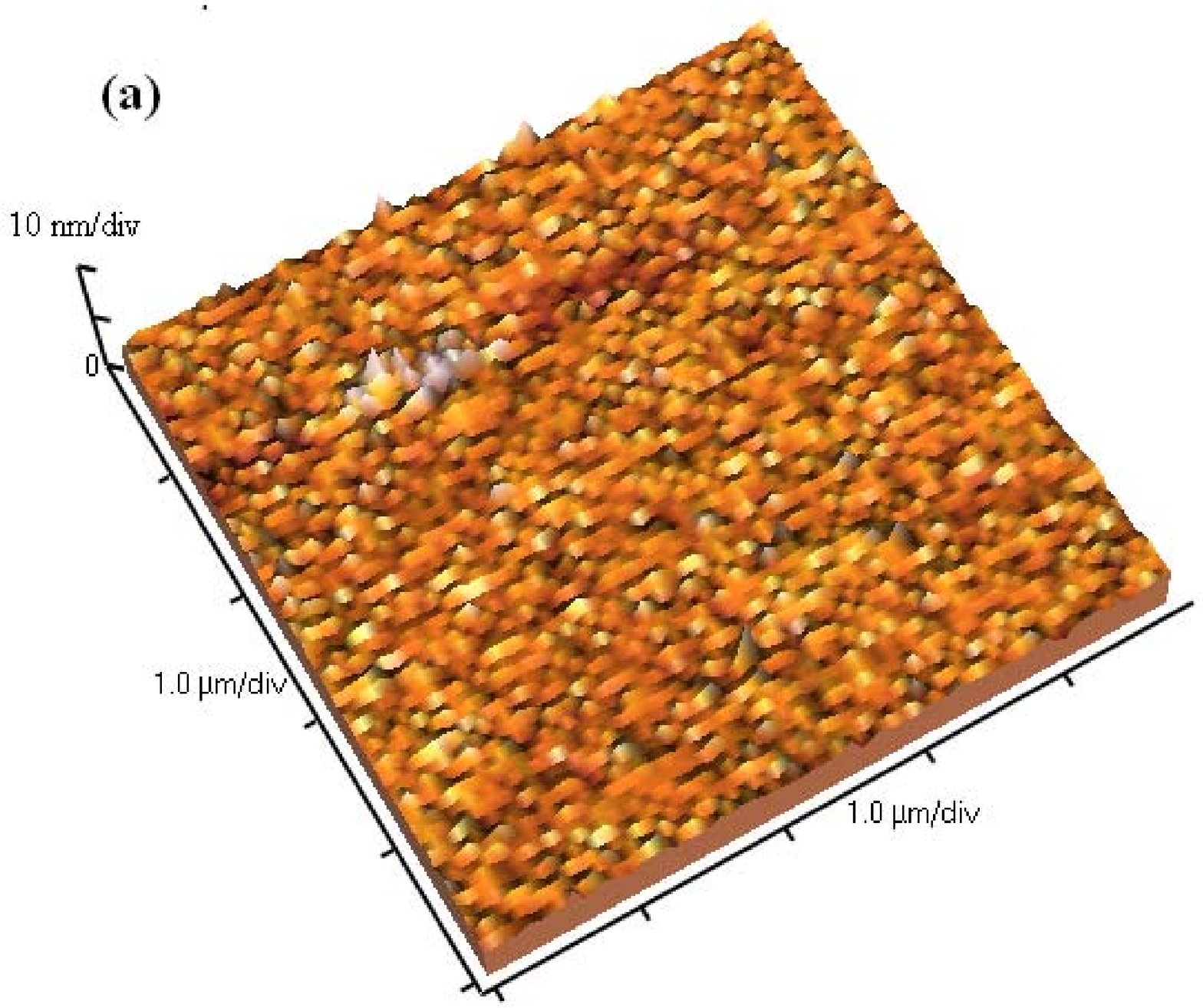}
\includegraphics[width=8truecm]{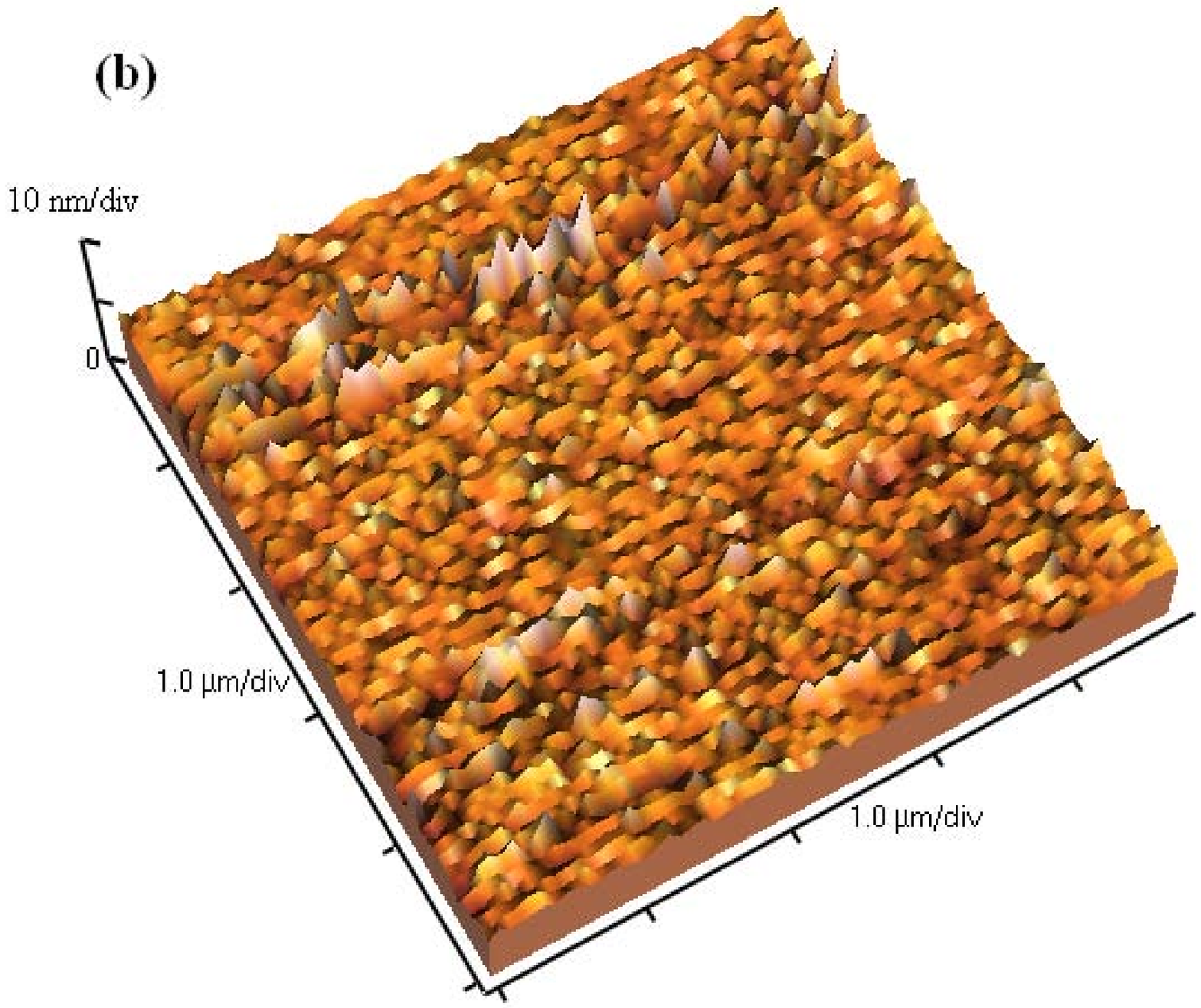}
\includegraphics[width=8truecm]{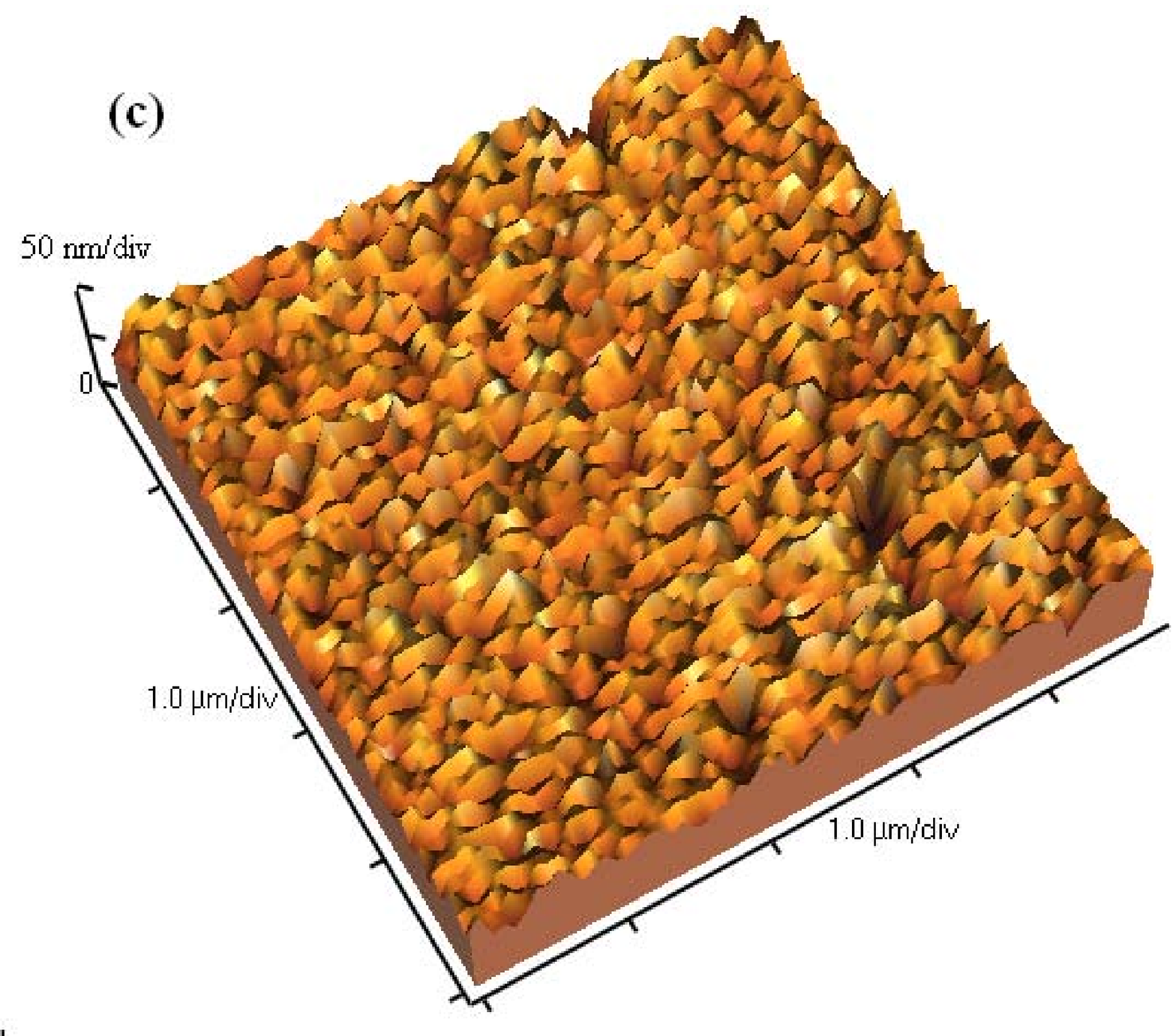}
\end{figure}
\begin{figure} 
\includegraphics[width=8.5truecm]{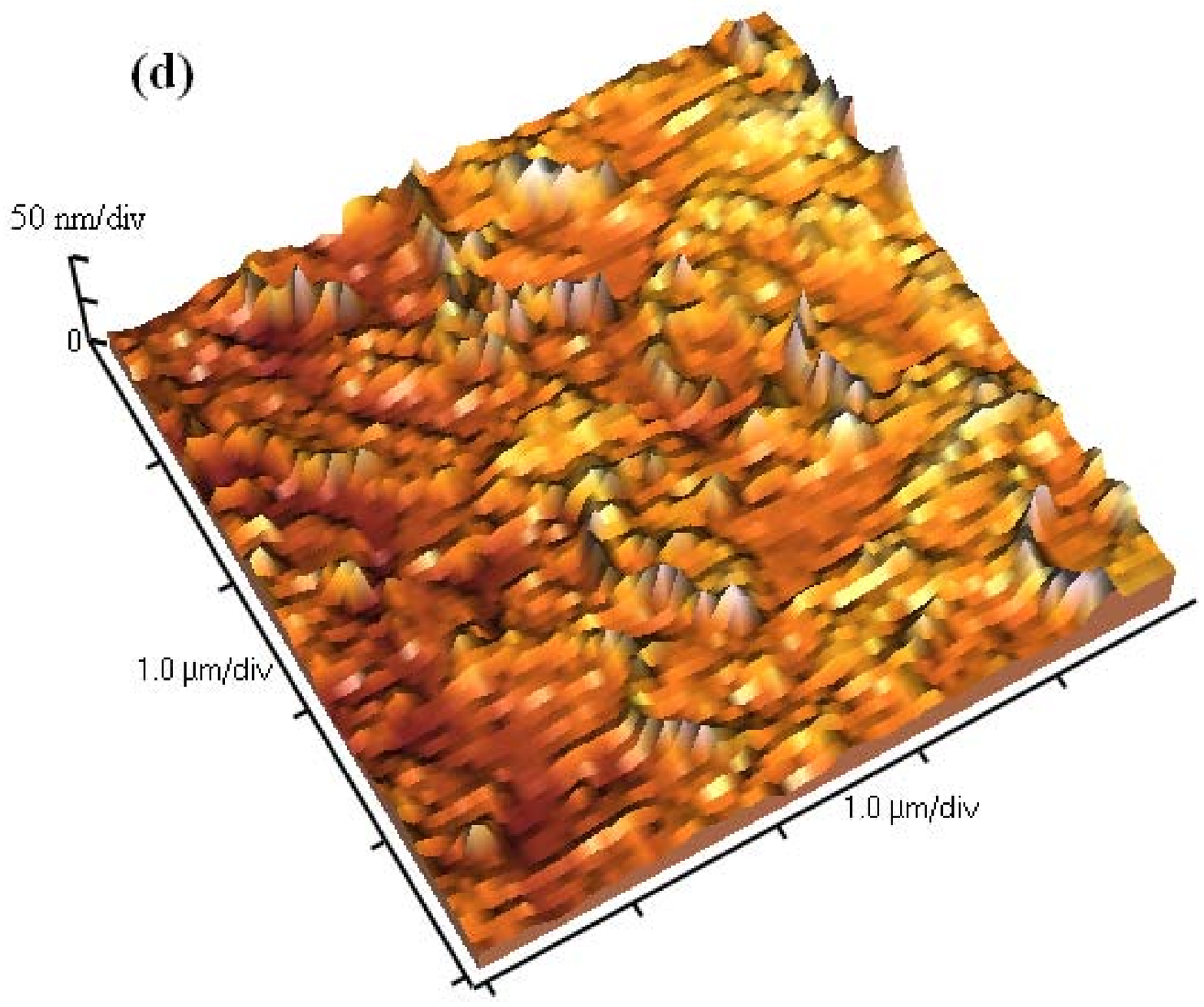}
\includegraphics[width=8.5truecm]{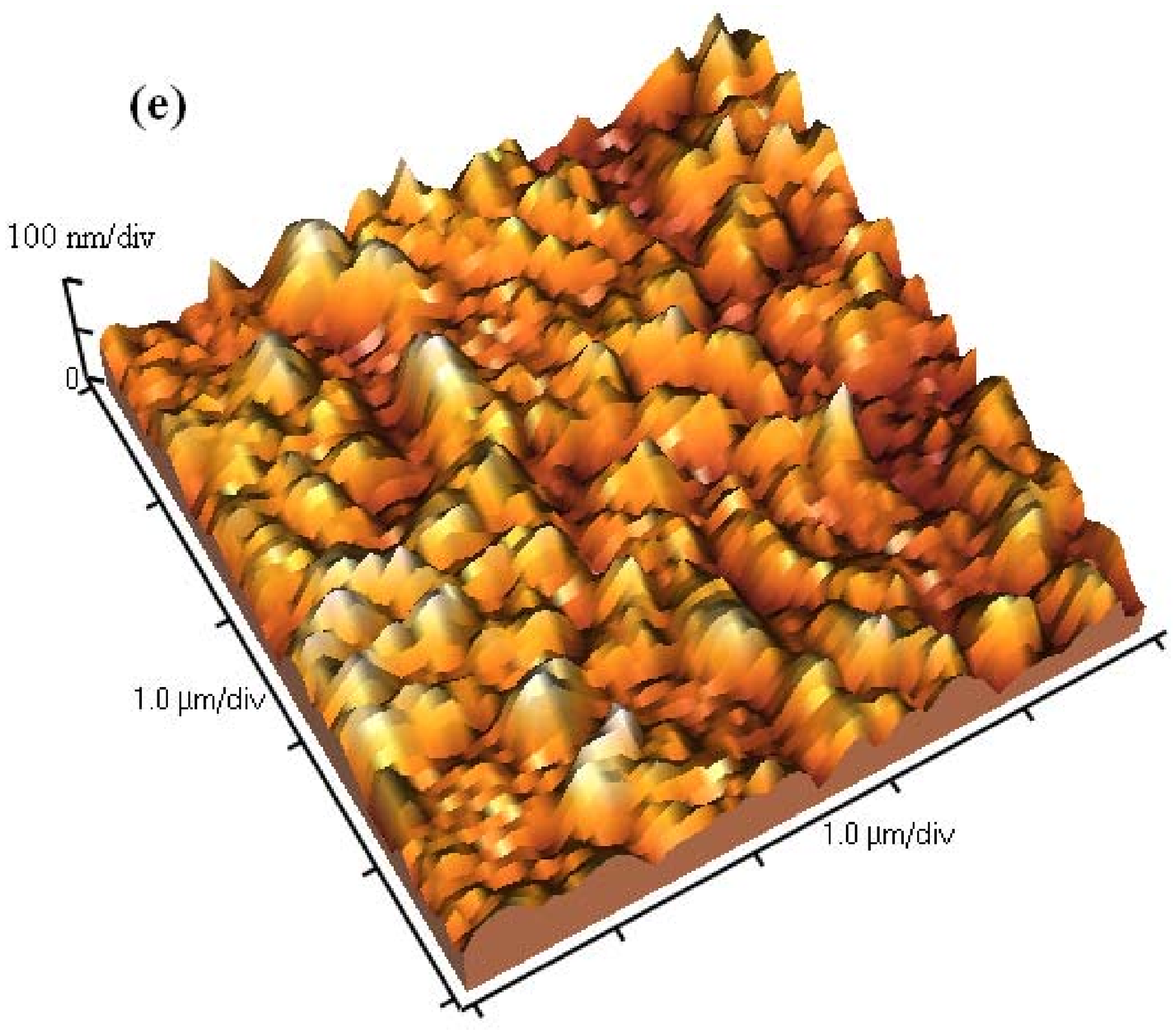}
\includegraphics[width=8.5truecm]{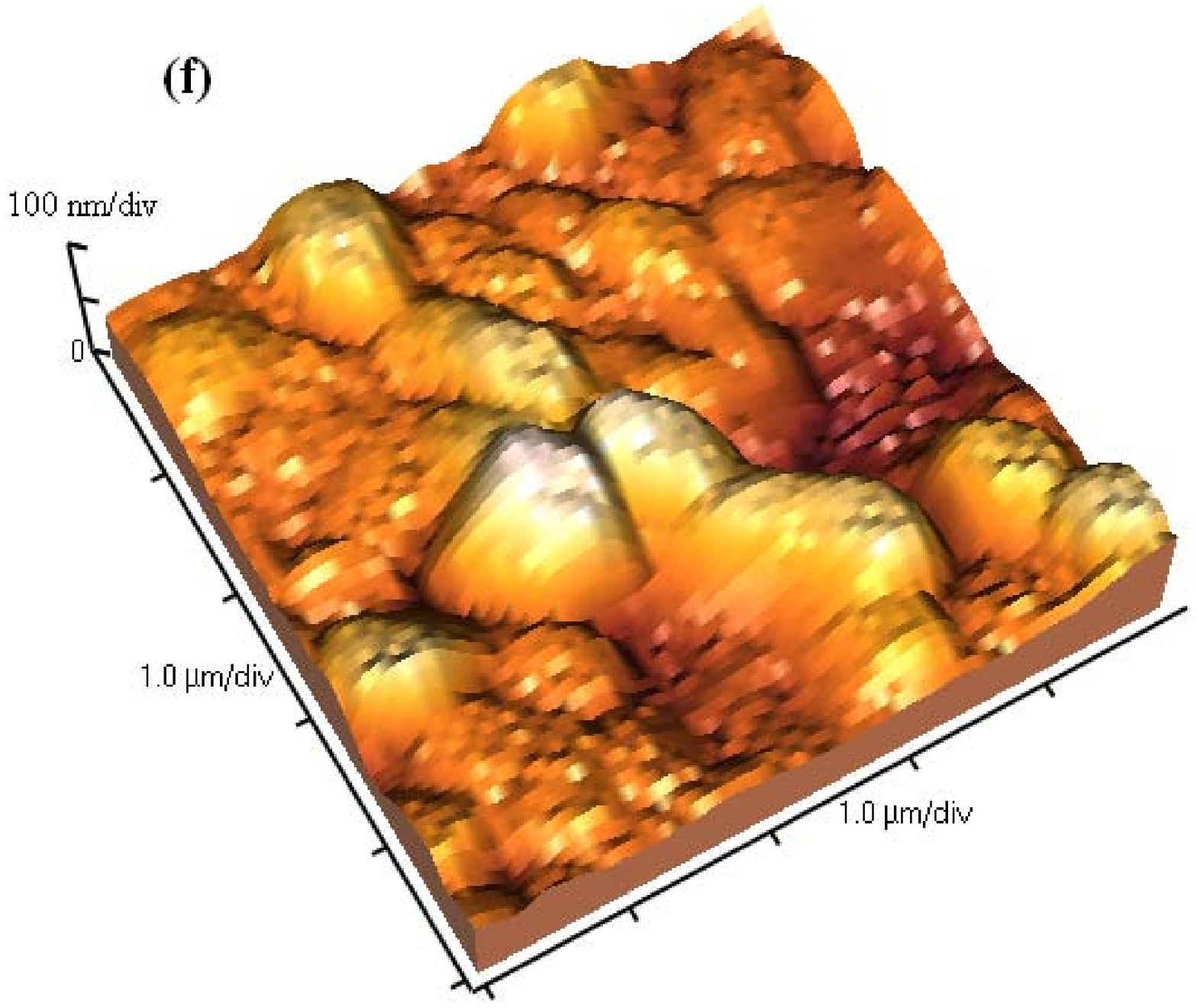}
\narrowtext \caption{ AFM images
of $WO_3$ thin films annealed at various temperatures a) 200, b)
300, c) 350, d) 400, e) 450 and f) 500$^oC$, respectively. }
\end{figure}

The optical gap ($E_g$) was evaluated from the absorption
coefficient ($\alpha$) using the standard relation: $(\alpha h
\nu)^{1/\eta}=A(h \nu-E_g)$, in which $\eta$ depends on the kind
of optical transition in semiconductors, and $\alpha$ was
determined near the absorption edge using the simple relation:
$\alpha=ln[(1-R)^2/T]/d$ ,where $d$ is thickness of the film. More
useful explanation about the optical band gap calculation
reported in \cite{Moshfegh}. The relationship between the optical
band gap energy and annealing temperature for $WO_3$ thin films
has been shown in Fig. 2b. As can be seen from it, the optical
band gap for the "as deposited" $WO_3$ evaluated 3.4 $eV$.
Amorphous structure of the "as deposited" $WO_3$ causes to $E_g$
is bigger than 2.7 $eV$. After annealing samples at 200 and $300
^oC$, the optical band gap decreased slightly about 0.1 $eV$
which can be related to condensation of the films. But the
optical band gap of the $WO_3$ annealed at $400^oC$ reduced to 3.1
$eV$ due to crystallization of the film. This reduction continues
to 2.5 $eV$ for the sample annealed at $500^oC$. Reason of the
$E_g$ becomes smaller than 2.7 $eV$ is oxygen vacancy at this
temperature as was seen in Fig 1b. It is to note that for
evaporated $WO_3$ films one has found $2.7 < E_g < 3.5$  $eV$ \cite{Granqvist}.\\

\subsection{AFM Analysis}

To study the effect of the annealing process on the surface
morphology of the films, we have shown AFM images of the $WO_3$
surfaces annealed at the different temperatures : 200, 300, 350,
400, 450, 500$^oC$ in Figure 1. As can be seen from Fig. 1, for the
annealed film at 200$^oC$, it seems that the surface morphology of
the film is relatively the same with a smooth surface, amorphous
structure and nanometric grain size, as also reported by other
investigators for $WO_3$ films \cite{Antonik,Mohammad}. We have also
observed similar image for the "as deposited" $WO_3$ which is not
shown here. For $WO_3$ thin films, increasing annealing temperature
to $350^oC$ did not significantly effect on surface parameters
because it is low temperature for crystallization of $WO_3$
\cite{Granqvist}. But at higher annealing temperatures 400, 450 and
500$^oC$, surface grain size and roughness begin to increase. The
more precise analysis of these surfaces are given in the next
section.


\section{Statistical quantities}

\subsection{Roughness Analysis}

It is also known that to derive the quantitative information of the
surface morphology one may consider a sample of size $L$ and define
the mean height of growing film $\overline{h}$ and its roughness
$\sigma$ by:
\begin{equation}\label{w}
\sigma(L,t)=(\langle (h-\overline{h})^2\rangle)^{1/2}
\end{equation}
where $t$ is growing time and $\langle\cdots\rangle$ denotes an
averaging over different samples, respectively. Moreover, growing
time is a factor which can be applied to control the surface
roughness of thin films.

Let us now calculate the roughness exponent of the growing surface.
Starting from a flat interface (one of the possible initial
conditions), it is conjectured that a scaling of lenght by factor
$b$ and of time by factor $b^z$ ($z$ is the dynamical scaling
exponent), rescales the roughness $\sigma$ by factor $b^{\chi}$ as
follows \cite{Barabasi}:
\begin{equation}\label{scaling}
\sigma(bL,b^zt)=b^{\alpha}\sigma(L,t)
\end{equation}
which implies that
\begin{equation}
\sigma(L,t)=L^{\alpha}f(t/L^z).
\end{equation}
For large $t$ and fixed $L$ $(i.e.  x=t / L^z \rightarrow \infty)$
$\sigma$ saturate.  However, for fixed and large $L$ and $t\ll L^z$,
one expects that correlations of the height fluctuations are set up
only within a distance $t^{1/z}$ and thus must be independent of
$L$. This implies that for $x \ll 1$, $f(x)\sim
x^{\beta}g'(\lambda)$ with $\beta=\alpha / z$. Thus, dynamic scaling
postulates that
\begin{eqnarray}
\sigma(L,t)=
\left\{%
\begin{array}{ll}
   t^{\beta}, & \hbox{t$\ll L^z$;}\\
   L^{\alpha}, & \hbox{t$\gg L^{z}$}. \\
\end{array}%
\right.
\end{eqnarray}
The roughness exponent $\alpha$ and the dynamic exponent $\beta$
characterize the self-affine geometry of the surface and its
dynamics, respectively. In the present work, we see the surfaces
at the limit $t \rightarrow \infty$ and so we will only obtain
the $\alpha$ exponent.

The common procedure to measure the roughness exponent of a rough
surface is use of the surface structure function depending on the
length scale $l$ which is defined by :
\begin{eqnarray}\label{Structure}
S^{2}(l)=\langle|h(x+l)-h(x)|^2\rangle.
\end{eqnarray}
It is equivalent to the statistics of height-height correlation
function $C(l)$ for stationary surfaces, i.e.
$S^{2}(l)=2\sigma^2(1-C(l))$. The second order structure function
$S^2(l)$, scales with $l$ as $ l^{2\alpha}$.

\subsection{Level Crossing Analysis}

Let $\nu_{\alpha}^{+}$ denotes the number of positive slope crossing
of $h(x)- \bar h = \alpha$ for interval L.

Since the process is homogeneous, if we take a second time interval
of $L$ immediately following the first we shall obtain the same
result, and for two intervals together we shall therefore obtain
\cite{Tabar1}:
\begin{equation}
N_{\alpha}^{+}(2L)=2N_{\alpha}^{+}(L),
\end{equation}
from which it follows that, for a homogeneous process, the average
number of crossing is proportional to the interval $L$. Hence
\begin{equation}
N_{\alpha}^{+}(L)\propto L,
\end{equation}
or
\begin{equation}\label{n}
N_{\alpha}^{+}(L)=\nu^{+}_{\alpha} L,
\end{equation}
where $\nu_{\alpha}^{+}$ is the average frequency of positive slope
crossing of the level $h(x) - \bar h =\alpha$. We now consider how
the frequency parameter $\nu_{\alpha}^{+}$ can be deduced from the
underlying probability distributions for $h(x) - \bar h$.\\ Consider
a small length scale $\delta x$ of a typical sample function. Since
we are assuming that the process $h(x)-\bar h$ is a smooth function
of $x$, with no sudden ups and downs, if $\delta x$ is small enough,
the sample can only cross $h(x)-\bar h=\alpha$ with positive slope
if $h(x)-\bar h < \alpha$ at the beginning of the interval $L$.
Furthermore, there is a minimum slope at $x$ if the level $h(x)-
\bar h = \alpha$ is to be crossed in interval $\Delta x$ depending
on the value of $h(x)-\bar h$ at position $x$. So there will be a
positive crossing of $h(x)-\bar h =\alpha$ in the next interval
$\Delta x$ if, at $x$,

\begin{equation}
h(x)- \bar h < \alpha \hspace{.6cm} and \hspace {.6cm}
\frac{\Delta\left[h(x)-\bar h\right]}{\Delta x}>
\frac{\alpha-\left[h(x) - \bar h\right] }{\Delta x}.
\end{equation}

As shown in \cite{Tabar1}, the frequency $\nu_{\alpha}^{+}$ can be
written in terms of joint PDF (probability distribution function )
of $p(\alpha,{y}^{\prime})$ as follows:
\begin{equation}\label{level}
\nu_{\alpha}^{+}=\int_{0}^{\infty}p(\alpha,{y}^{\prime}){y}^{\prime}d{y}
^{\prime}.
\end{equation}
and then the quantity $N_{tot}^{+}$ which is defined as:
 \begin{equation}\label{ntq}
N_{tot}^{+}=\int_{-\infty}^{+\infty}\nu_{\alpha}^{+} |\alpha -
\bar{\alpha}| d \alpha.
\end{equation}
will measure the total number of crossing the surface with
positive slope. So, the $N_{tot}^{+}$  and square area of growing
surface are in the same order. Concerning this, it can be
utilized as another quantity to study further the roughness of a
surface \cite{Tabar}.
\begin{figure}
\includegraphics[width=10.4truecm]{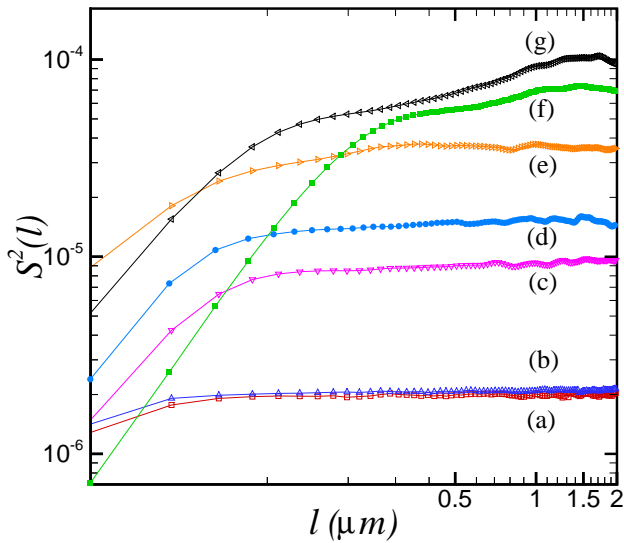}
 \narrowtext \caption{ Log-Log
plot of the selection structure function of various annealed
temperature: a) 27, b) 200, c) 300, d) 350, e) 400, f) 450, g)
500$^oC$.}
\end{figure}

\section{Results and Discussion}

Thin films of $WO_3$ were deposited by using thermal evaporation
method and then surface micrographs of $WO_3$ samples were obtained
by AFM technique after annealed at different temperatures (Fig.3).\\
These micrographs were then analyzed using methods from stochastic
data analysis have introduced in the last section. Figure 3 shows
AFM images of $WO_3$ thin films annealed at 200 ,300 ,350 ,400 ,450,
and 500$^oC$. The "as deposited" and annealed sample at 200$^oC$
(Fig. 3a) have columnar structure, indicating that up to 200$^oC$ no
significant changes in the microstructure occurs. However, at higher
temperatures (figs. 3b-3f) we have observed increased grain size and
rougher surface. Specifically at 500$^oC$ (Fig. 3f) we observe stark
changes in the micrograph which is accompanied by composition
changes in the surface. This can be related to the phase transition
to Magneli phase e.g. $WO_{3-x}$ in the annealing process
\cite{Mohammad}. This is also confirmed by our XPS and UV-visible
spectrophotometry analysis
(Sec. II). These are shown the significant formation of $W^{5+}$ state in the surface at 500$^oC$.\\

Also our analysis shows that below 400$^oC$ the surfaces are in
amorphous phase with the same behavior for all scales, but as soon
as the crystalline phase appears the system behaves differently
which diagnostics at small and large scale for temperatures above
400$^oC$. By using parameters of the analytical
method given here, these transitions can be quantified.\\
\begin{figure}
\includegraphics[width=9.4truecm]{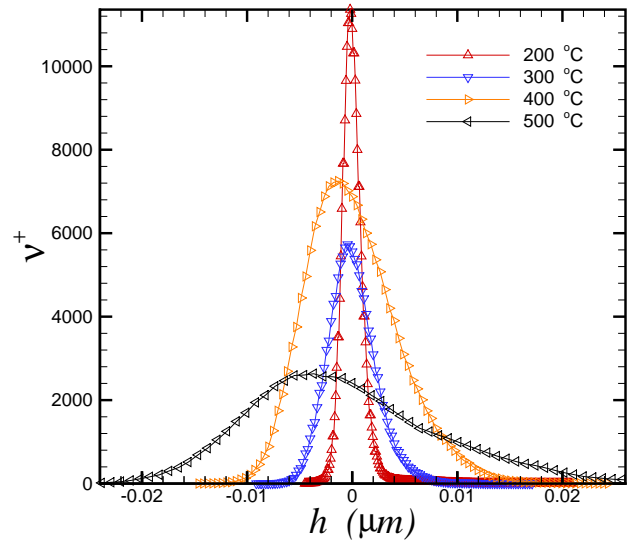}
 \narrowtext \caption{ The average
frequency $\nu_{\alpha}^{+}$ as a function of height $h$}
\end{figure}

Now, we will use the statistical parameters introduced in the last
section and will obtain some quantitative information about the
effect of annealing temperature on the surface topography of the
$WO_3$ samples.\\

The structure function $S^2(l)$ as defined in Eq.(5) can be used to
quantify the topology of a rough surface. The structure function
$S^2(l)$ is plotted against the length scale of the sample in Fig.4
. The saturated $S^2(l)$ is an indication of the surface roughness,
as $2\sigma^2$. The most obvious observation indicates that
roughness is raised with increasing annealing temperature. Roughness
has a minimum of $0.91 nm$ at 27 and 200$^oC$ and a maximum of $48
nm $ at 500$^oC$. This is because higher temperatures create higher
peaks (i.e. peaks with more deviations from the average) . All
exponents which is derivable from $S^2(l)$ have been summarized and
given in Table I.\\

As depicted in Fig.4 , the structure function $S^2(l)$, has a
different behavior in the various temperatures. So that, in the
annealing temperature range 27-350$^oC$ it has a typical behavior in
all scales, but in the higher temperature range 400-500$^oC$ its
behavior is different in the small and large scales. In the other
words, the phase transition is occurred at 400$^oC$, because for
higher temperatures, there are two sets of roughness parameters
needed to simulate the surface morphology. It can be related to the
phase transition in the structure of the surface from amorphous to
crystalline phase has been yielded from the band gap energy (see the
section II.B).\\

The slope of each $S^2(l)$ curves at the small and large scales
yields the roughness exponents $\alpha$ and $\alpha'$ of the
corresponding surface. Hence, it is seen that the mono roughness
exponent increases with the addition of annealing temperature up to
400$^oC$. In the higher temperatures, we have obtained two roughness
exponents( $\alpha$-$\alpha'$) equal to the 0.40-0.14, 0.71-0.20,
and 0.69-0.24 for temperatures 400, 450 and 500$^oC$, respectively.
Difference in the $\alpha$ values, in these temperatures, are in
agreement with changes of correlation length.
Where the correlation length, is the distance at which the structure function behaves differently.\\

The range of the scaling upon correlation length listed in the forth
column in table I. The value of $C_{s}^{\ast}$ denotes the
correlation length at small scales and $C_{l}^{\ast}$ for large
scales. The higher $C^{\ast}$ value represents a smoother surface
(as we expected from Fig.3). The correlation length obtained from
the structure function is also a measure of minimum lateral size of
surface features at each annealing temperature. \\

\begin{figure}
\includegraphics[width=8.4truecm]{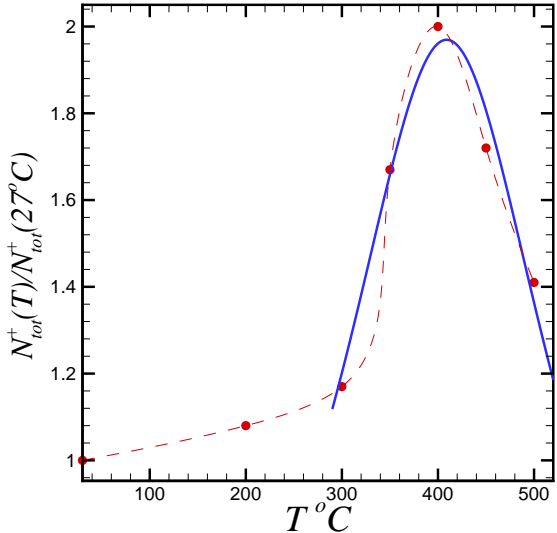}
 \narrowtext \caption{ The normalized $N^+_{tot}$ behavior
 as a function of annealing temperature. The solid line is plotted according to Eq.(12) around 400$^oC$ .}
\end{figure}

The another important $WO_3$ film parameter is the effective area of
the sample which has an important role in the gas sensitivity of
$WO_3$ surfaces. To obtain a measure for this, we utilize the level
crossing analysis. As shown in Fig.5, the average frequency
$\nu_{\alpha}^{+}$ as a function of height $h$, is plotted for the
various annealing temperatures. The broad curves indicate the higher
magnitude of height fluctuations around the average, and sharp
curves show that the most of fluctuations are around the height
average. This conclusion is in the correspondence with the results
obtained from Fig.3.

According to the Eq.(11) $N_{\alpha}^{+}$ i.e. The total number of
the crossing surface with positive slope is proportional to the
square of area of the growing surface. To obtain the optimum value
of the effective area, we have calculated the ratio of effective
areas with respect to the area of the "as deposited" surface
(27$^oC$).
\begin{table}[htb]
\begin{center}\label{Tb2}
\caption{The Roughness exponent, roughness, correlation length and
effective area relative to the "as deposited" sample area (27$^oC$).}
\begin{tabular}{|c|c|c|c|c|}
T $[^oC]$ &$\alpha-\alpha'$&$\sigma$$[n
m]$&$C_{s}^{\ast}$-$C_{l}^{\ast}$$[nm]$& $N^{+}/N^{+}(27^oC)$
\\\hline
$27$  &$0.15-none$& $0.91$ & $60-none$&$1.00$
\\\hline
$200$ &$0.15-none$& $0.98$ & $60-none$&$1.08\pm0.02$
\\\hline
$300$ &$0.61-none$& $2.20$ & $100-none$&$1.17\pm0.02$
\\\hline
$350$ &$0.62-none$& $11.50$ & $100-none$&$1.67\pm0.02$
\\\hline
$400$ &$0.40-0.15$& $17.00$ & $100-300$&$2.00\pm0.02$
\\\hline
$450$ &$0.71-0.20$& $30.00$ & $400-1000$&$1.72\pm0.02$
\\\hline
$500$ &$0.69-0.24$& $48.00$ & $200-1400$&$1.41\pm0.02$
\end{tabular}
\end{center}
\end{table}
The values are given in the last column in Table I. It means,
although the roughness increases by the annealing temperature but
the effective square area of the rough surface has a maximum value
of $N_{tot}^{+}$ \cite{Tabar}.\\
For more clarity, we have calculated the temperature dependence of
normalized $N_{\alpha}^{+}$ numerically (Fig.6) around 400$^oC$, and
we have obtained the three following functions for this quantity :\\
\begin{eqnarray}
N^+_{tot}(T)=(5.0718-0.0223\times T+2.72\times10^{-5}\times
T^2)^{-1}
\end{eqnarray}
\begin{eqnarray}
ln(N^+_{tot}(T))=-6.3632+0.0344\times T-4.20\times10^{-5}\times T^2
\end{eqnarray}
\begin{eqnarray}
N^+_{tot}(T)=-8.7057+0.0520\times T-6.37\times10^{-5}\times T^2
\end{eqnarray}

According to this figure, the maximum value of the effective area is
at 400$^oC$ (with respect to its value at 27$^oC$) with the relative
value equal to 2.00. Thus, applying this analysis easily shows that
if one follows the condition which the effective area as an
important parameter in the gas sensitivity of $WO_3$ surfaces is
optimum and furthermore, the film composition has not been changed
(e.g. The Magneli phase transition has not been
occurred), should choose the annealed surface at 400$^oC$ for better performance.\\


\section{Conclusions}

We have investigated the role of annealing temperature, as an
external parameter, to control the statistical properties of a rough
$WO_3$ surface. The AFM microstructure of the surfaces is just
needed to apply in our analysis. We have computed the statistical
quantities such as roughness exponent, roughness and lateral size of
surface features of the "as deposited" and annealed surfaces at 200,
300, 350, 400, 450, and 500$^oC$, using the structure function. We
have seen a phase transition at 400$^oC$, because for higher
temperatures there are two sets of roughness parameters, due to
structural changes from amorphous to the crystalline phase.
Moreover, using the level crossing analysis we have obtained an
optimum annealing temperature, 400$^oC$ in which the surface of the
$WO_3$ has maximum value about twice relative to the "as deposited"
film without any changes in the film composition that may increase
surface reaction of the $WO_3$ film as the gas sensor or
photo-catalyst.

\section{Acknowledgment}

GRJ and AAS would like to thank S.M.Fazeli  for his useful comments
and especially M.R.Rahimitabar for his useful lectures on
"stochastic data analysis". AZM would like to acknowledge research
council of Sharif University of Technology for financial support of
the work.

\end{document}